\documentstyle[aps,prl,multicol,epsfig,fancyheadings]{revtex}
\renewcommand{\[}{\begin{equation}}
\renewcommand{\]}{\end{equation}}
\newcommand{\equ}[1]{Eq.~(\ref{#1})}
\newcommand{\eqs}[2]{Eqs.~(\ref{#1}) and (\ref{#2})}

\newcommand{\ei}[1]{{\rm e}^{i #1}}

\newcommand{\kk}{\mbox{\boldmath$\kappa$}}
\newcommand{\intr}{\int \! d{\bf r} \;}
\newcommand{\da}{\partial_\alpha}
\newcommand{\db}{\partial_\beta}

\begin{document}

\draft
\twocolumn[\hsize\textwidth\columnwidth\hsize\csname@twocolumnfalse\endcsname

\title{Electron Localization in the Quantum-Hall Regime}

\author{Raffaele Resta}

\address{INFM DEMOCRITOS National Simulation Center, via Beirut 2,
I--34014 Trieste, Italy}
\address{and Dipartimento di Fisica Teorica, Universit\`a di
Trieste,\\ Strada Costiera 11, I--34014 Trieste, Italy}

\date{}

\maketitle

\begin{abstract} The theory of the insulating state discriminates between
insulators and metals by means of a localization tensor, which is finite in
insulators and divergent in metals. In absence of time-reversal symmetry,
this same tensor acquires an offdiagonal imaginary part, proportional to the
dc transverse conductivity, leading to quantization of the latter in
two-dimensional systems. I provide evidence that electron localization---in
the above sense---is the common {\it cause} for both vanishing of the dc
longitudinal conductivity and quantization of the transverse one in
quantum-Hall fluids.  \end{abstract}

\bigskip\bigskip

]
\narrowtext

W. Kohn showed in 1964 that the insulating state of matter reflects a peculiar
organization of the electrons in their ground state: the cause for the
insulating behavior is electron localization~\cite{Kohn64,Kohn68}.  Such
localization, however, manifests itself in a very subtle way, fully elucidated
much later. In 1999 Resta and Sorella~\cite{rap107} defined a tensor which
provides a quantitative measure of Kohn's localization, and has a common root
with the modern theory of polarization~\cite{modern,rap100,Martin22}. This
``localization tensor'' is an intensive property characterizing the ground
wavefunction as a whole: it is finite in any insulator and divergent in any
metal. A further advance on this line was provided in 2000 by Souza, Wilkens,
and Martin~\cite{Souza00}. I am going to refer to these results altogether as to
the ``theory of the insulating state'' (TIS)~\cite{rap_a23}: so far, it
has only considered time-reversal-invariant systems. I show here that, in
absence of time-reversal symmetry, the TIS localization
tensor~\cite{rap107,Souza00,rap_a23} is naturally endowed with a nonvanishing
imaginary part. For a two-dimensional system, the imaginary part is quantized
whenever the real part is non divergent, and is proportional to dc transverse
conductivity. I show here that the theory of the quantum-Hall effect
(particularly in the formulation of Niu, Thouless, and Wu~\cite{Niu85}) has a
very direct---and previously unsuspected---relationship to TIS, and in fact can
be regarded as a consequence of the latter. In order to predict whether the dc
transverse conductivity of any two-dimensional many-electron system is
quantized, it is enough to inspect electron localization in the ground state: this
is the major result of the present Letter.

Phenomenologically, an insulating material is characterized by vanishing dc
longitudinal conductivity. In this sense, an electron fluid in the quantum-Hall
regime is in fact an insulator, independently of what establishes such
regime (e.g. disorder). According to TIS the many-body wavefunction is then {\it
localized}. From the present viewpoint, electron localization is the common {\it
cause} for both vanishing of the longitudinal dc conductivity and quantization
of the transverse one; the two features stem here from the same formalism. The
present view may appear at odds with the established one, which in the
quantum-Hall regime focusses on the extended states more than on the localized
ones~\cite{Yoshioka}; but it is worth stressing that the TIS localization tensor
is a global geometric property characterizing the ground wavefunction, {\it not}
the individual one-electron states. 

In the final part of this Letter I also show how TIS works for noninteracting
electrons in the lowest Landau level. While disorder is obviously essential  for
producing a quantum-Hall fluid, a flat substrate potential is used here to
provide analytical results. At complete filling the (real) trace of the
localization tensor is shown to be equal to the squared magnetic length, while
the (imaginary) antisymmetric part of the same tensor provides the Hall
conductivity; at fractional filling the real part diverges while the imaginary
part is ill defined. This confirms our main message: inspecting the ground-state
localization is enough to predict quantization of transverse conductivity.

The TIS localization tensor~\cite{rap107}, also known as second cumulant
moment $\langle r_\alpha r_\beta \rangle_{\rm c}$ of the electron
distribution~\cite{Souza00,rap_a23}, is an intensive property having the
dimensions of a squared length, and whose only ingredient is the many-body
ground wavefunction $| \Psi_0 \rangle$. In the cases dealt with so far,
periodic boundary conditions were adopted; these are easily modified to
accomodate a macroscopic magnetic field~\cite{Niu85}. If $| \Psi_0 \rangle$
is an $N$-electron wavefunction periodic with period $L$ in all Cartesian
coordinates $r_{j,\alpha}$ separately, we define $\kk_\alpha = (2\pi/L) {\bf
e}_\alpha$, where ${\bf e}_\alpha$ is a unit vector along $\alpha$, and \[ |
\Psi_0(0) \rangle = | \Psi_0 \rangle ; \quad | \Psi_0(\kk_\alpha) \rangle =
\ei{\frac{2\pi}{L} \sum_{j=1}^N r_{j,\alpha}} | \Psi_0 \rangle.  \label{kk}
\] According to TIS, the localization tensor
is~\cite{rap107,Souza00,rap_a23}: \[ \langle r_\alpha r_\beta \rangle_{\rm c}
= \frac{L^2}{4 \pi^2 N} \ln \frac{ \langle \Psi_0(\kk_\alpha) |
\Psi_0(\kk_\beta) \rangle}{\langle \Psi_0(\kk_\alpha) | \Psi_0(0) \rangle
\langle \Psi_0(0) | \Psi_0(\kk_\beta) \rangle} , \label{defd} \] where the
thermodynamic limit is understood. In the existing literature time-reversal
symmetry is assumed: the tensor is then real.  When time-reversal invariance
is absent, this same tensor is endowed with an off-diagonal imaginary part,
which---as shown below---is particularly relevant for two-dimensional systems.

As first shown by Souza, Wilkens, and Martin~\cite{Souza00} by means of a
fluctuation-dissipation theorem, the real part of the localization tensor is
related to a frequency integral of the longitudinal conductivity, which is
finite in any insulator and divergent in any metal: \begin{eqnarray}
\int_0^\infty  \frac{d \omega}{\omega} \; \mbox{Re }
\sigma_{\alpha\alpha}(\omega) = \frac{\pi e^2 N}{\hbar L^d} \mbox{Re }\langle
r_\alpha^2 \rangle_{\rm c} \nonumber \\ = - \frac{e^2}{4 \pi \hbar L^{d-2}}
\ln | \langle \Psi_0(0) | \Psi_0(\kk_\alpha) \rangle|^2 .  \label{swm}
\end{eqnarray} I am going to extend this result, in order to address the
offdiagonal imaginary part of the localization tensor as well, and
additionally to consider cases where a macroscopic magnetic field is present.
Specializing from now on to a two-dimensional system, we notice that
\equ{swm} is size-invariant in form.

I assume the system as isotropic in the $xy$ plane, with a magnetic field $B$
along $z$. Therefore $\sigma_{11} = \sigma_{22}$, while the offdiagonal
element is purely antisymmetric: $\sigma_{12} = - \sigma_{21}$.  The Kubo
formula for the conductivity tensor is: \begin{eqnarray} \sigma_{\alpha\beta}
(\omega) = \frac{i e^2}{\hbar \omega L^2} \lim_{\eta \rightarrow 0+} {\sum_{n
\neq 0}}' \left( \frac{\langle \Psi_0 | \hat{v}_\alpha | \Psi_n \rangle
\langle \Psi_n | \hat{v}_\beta | \Psi_0 \rangle }{\omega - \omega_{0n} + i
\eta} \right.  \nonumber \\ - \left. \frac{\langle \Psi_0 | \hat{v}_\beta |
\Psi_n \rangle \langle \Psi_n | \hat{v}_\alpha | \Psi_0 \rangle }{\omega +
\omega_{0n} + i \eta} \right) , \label{kubo} \end{eqnarray} where
$\omega_{0n} = (E_n - E_0)/\hbar$ are the excitation frequencies. I then
focus on the the two quantities: \begin{eqnarray} \int_0^\infty \frac{d
\omega}{\omega} \; \mbox{Re}&&\!\!\!\! \sigma_{11}(\omega) \nonumber \\ & = &
\frac{\pi e^2}{\hbar L^2} \mbox{Re } \sum_{n \neq 0} \frac{ \langle \Psi_0 |
\hat{v}_1 | \Psi_n \rangle \langle \Psi_n | \hat{v}_1 | \Psi_0 \rangle
}{\omega_{0n}^2} \label{sum1} \\ \mbox{Re } \sigma_{12}(0) & = & \frac{2
e^2}{\hbar L^2} \mbox{Im } \sum_{n \neq 0} \frac{\langle \Psi_0 | \hat{v}_1 |
\Psi_n \rangle \langle \Psi_n | \hat{v}_2 | \Psi_0 \rangle }{\omega_{0n}^2} ,
\label{sum2} \end{eqnarray} where the r.h.s. member are written as to
emphasize the common structure. Notice that we have taken the limit $\eta
\rightarrow 0$ at finite $L$. In transport theory the interest is in
evaluating $\sigma$ as a continuous function of $\omega$, by smoothing the
singularities in \equ{kubo}: this can be done by keeping the ``dissipation''
$\eta$ finite while performing the thermodynamic limit
first~\cite{Akkermans97}. The order of the two limits is irrelevant here,
since \equ{sum1} is an {\it integrated} property, and \equ{sum2} is
dissipationless.

In order to transform the sum over the excited states into a
ground-state property, it is expedient to consider the many-body Hamiltonian
with a ``twist'' (or ``flux'') \[ \hat{H}({\bf k}) = \frac{1}{2m} \sum_{i=1}^N
\left({\bf p}_i - \hbar {\bf k} + \frac{e}{c} {\bf A} \right)^2 + \hat{V} ,
\label{many} \] where $\hat{V}$ comprises the one-body substrate potential and
the electron-electron interaction.  We indicate the ground state of
\equ{many} as $| \Psi({\bf k}) \rangle$, with $| \Psi(0) \rangle = | \Psi_0
\rangle$;  straightforward perturbation theory yields \[ | \Psi({\bf k}) \rangle
\simeq | \Psi_0 \rangle + {\bf k} \cdot \sum_{n\neq 0} | \Psi_n \rangle
\frac{\langle \Psi_n | \hat{\bf v} | \Psi_0 \rangle}{\omega_{0n}} , \] \[
\langle \Psi_n | \da \Psi(0) \rangle = \langle \Psi_n | \hat{v}_\alpha | \Psi_0
\rangle / \omega_{0n} , \qquad n \neq 0 \] where the velocity operator is
$\hat{\bf v} = {\bf \nabla}_{{\bf k}} \hat{H}({\bf k}) / \hbar$, and $\da \equiv
\partial/\partial k_\alpha$.  Strictly speaking, the perturbation expansion
holds for a conventional insulator where the Fermi gap does not vanish in the
thermodynamic limit. More generally, owing to \equ{sum1}, it also holds whenever
$\mbox{Re } \sigma_{\alpha\alpha}(\omega)$ goes to zero fast enough at small
$\omega$, i.e.  for any insulator~\cite{Souza00}.

The sum over excited states appearing in \eqs{sum1}{sum2} can then be
transformed into: \begin{eqnarray} & &\sum_{n \neq 0} \frac{\langle \Psi_0 |
\hat{v}_\alpha | \Psi_n \rangle \langle \Psi_n | \hat{v}_\beta | \Psi_0
\rangle }{\omega_{0n}^2} \label{PV} \\ &=& \langle \da
\Psi(0) | \db \Psi(0) \rangle - \langle \da \Psi(0) | \Psi(0) \rangle \langle
\Psi(0) | \db \Psi(0) \rangle \nonumber  \end{eqnarray} The real part of
\equ{PV} is the quantum metric tensor defined by Provost and
Vallee~\cite{Provost80}, evaluated at ${\bf k}=0$; the imaginary part is the
corresponding curvature (divided by two).

So far, we have specified neither the magnetic gauge nor the boundary
conditions. We choose the Landau gauge and the usual magnetic boundary
conditions~\cite{Niu85} for translations by $L$ of each
coordinate $x_i$ and $y_i$. These require the total flux $BL^2$ across the
system to be an integer number $N_s$ of flux quanta $\Phi_0 = hc/e$. At
filling $\nu$ the density is then \[ n_0 = \frac{\nu N_s}{L^2} = \frac{\nu}{2
\pi l^2}, \label{density} \] where $l = (\hbar c/e B)^{1/2}$ is the magnetic
length.

If the insulating ground state is nondegenerate at any ${\bf k}$, the
eigenstate $| \Psi({\bf k}) \rangle$ assumes a simple form whenever the ${\bf
k}$-coordinates are integer multiples of $2\pi/L$. For instance, if ${\bf k}$
coincides with one of the $\kk_\alpha$ vectors defined above, then $|
\Psi({\bf k}) \rangle$ coincides with \equ{kk} apart from a phase factor
which is irrelevant here: in fact the two wavefunctions obey the same
Schr\"odinger equation and the same magnetic boundary conditions. The case of
degenerate ground states has been considered as well~\cite{Aligia}. We then
discretize the derivatives in \equ{PV} using the special $\kk_\alpha$ vectors
of \equ{kk} and replacing $\langle \Psi_0(\kk_\alpha) | \Psi_0(0) \rangle
\simeq 1 + \ln \, \langle \Psi_0(\kk_\alpha) |  \Psi_0(0) \rangle$, as usual
when dealing with Berry phases~\cite{rap_a20}. The result is \begin{eqnarray}
& & \langle \da \Psi(0) | \db \Psi(0) \rangle - \langle \da
\Psi(0) | \Psi(0) \rangle \langle \Psi(0) | \db \Psi(0) \rangle
\label{discre} \\ &\simeq& \frac{L^2}{4\pi^2} \ln \frac{ \langle
\Psi_0(\kk_\alpha) | \Psi_0(\kk_\beta) \rangle}{\langle \Psi_0(\kk_\alpha) |
\Psi_0(0) \rangle \langle \Psi_0(0) | \Psi_0(\kk_\beta) \rangle} = N \langle
r_\alpha r_\beta \rangle_{\rm c}  \nonumber \end{eqnarray} Replacing the real
part of \equ{discre} into \eqs{sum1}{PV} one recovers the Souza-Wilkens-Martin
sum rule, \equ{swm}, which is nondivergent in the insulating case.

The imaginary part of \equ{discre} shares the same convergence properties as
the real one, after \equ{PV}; in the insulating case it takes the form of a
discrete Berry phase~\cite{rap_a20} over the three-point path in ${\bf
k}$-space from $0$ to $\kk_1$ to $\kk_2$ to $0$.  However, since Berry phases
are defined modulo $2 \pi$, this expression  does not provide a unique value.
The ambiguity is removed by replacing the Berry phase, i.e.  the loop
integral of the Berry connection, with the surface integral of the Berry
curvature. We therefore evaluate the imaginary part of \equ{PV} as
\begin{eqnarray} & & \mbox{Im } \langle \partial_1 \Psi(0) | \partial_2
\Psi(0) \rangle \nonumber \\ &=& \frac{L^2}{4\pi^2} \mbox{Im }
\int_0^{2\pi/L} \!\!\!\! dk_1 \int_0^{2\pi/L} \!\!\!\! dk_2 \;  \langle
\partial_1 \Psi({\bf k}) | \partial_2 \Psi({\bf k}) \rangle , \end{eqnarray}
in the limit of large $L$.  The dimensionless integral equals $- \pi C_1$,
where $C_1$, known as the first Chern number, is a topological
integer~\cite{Niu85,Kohmoto85,Thouless} characterizing the electron
distribution.  The imaginary part of the localization tensor is then
\begin{eqnarray} \mbox{Im } \langle xy \rangle_{\rm c} & = &\frac{1}{N}
\mbox{Im } \langle \partial_1 \Psi(0) | \partial_2 \Psi(0) \rangle \nonumber
\\ &=& - \frac{1}{4\pi} \frac{L^2}{N} C_1 = - \frac{l^2}{2\nu} C_1 . 
\label{chern} \end{eqnarray} Upon replacement of the previous expressions
into \equ{sum2} we retrieve the seminal result of Niu, Thouless, and
Wu~\cite{Niu85}: \[ \mbox{Re } \sigma_{12}(0) = - \frac{e^2}{h} C_1 . 
\label{NTW}\] This was originally obtained by an analysis of the Green
function, under the hypothesis that the system has a Fermi gap; in the
present approach the presence of a Fermi gap---possibly in the weak sense
outlined above---is a necessary and sufficient condition for the convergence
of \equ{PV} in the thermodynamic limit. But this property, belonging to the
{\it excitations} of the system, is transformed here into a pure {\it
ground-state} property, owing to a fluctuation-dissipation theorem. As far as
the longitudinal conductivity is concerned, a quantum-Hall fluid is no
different from any other insulator, and its wavefunction  is localized in the
sense of TIS~\cite{rap107,Souza00,rap_a23}. I have shown that, owing to such
localization,  any two-dimensional insulator may display a quantized
transverse conductance in absence of time-reversal symmetry (even in absence
of a macroscopic $B$ field~\cite{Haldane88}).

\equ{NTW} seems to legislate integer quantization of the Hall conductance in
all circumstances, contrary to experimental evidence. For fractional
fillings, Ref.~\cite{Niu85} assumes then a degenerate ground state, whose
different components are uncoupled and macroscopically separated. The
degeneracy problem has been thoroughly discussed in the literature (for a
review, see Ref.~\cite{Stone3}); the present Letter has nothing to add.

In the metallic case both sums in \eqs{sum1}{sum2} do not converge: the
former is positively divergent while the latter is indeterminate. Therefore
TIS formally defines the diagonal elements of the localization tensor as
infinite~\cite{rap107,Souza00,rap_a23} (delocalized ground wavefunction). 
The off-diagonal element $\langle xy \rangle_{\rm c}$ however, remains ill
defined, and the Kubo formula, \equ{sum2} is invalid. The transverse dc
conductivity is therefore not quantized as in \equ{NTW} and has to be
evaluated by different means, e.g. classically~\cite{Yoshioka}.

In the final part of this Letter we specialize to noninteracting electrons
and to the integer quantum-Hall effect. In the noninteracting case (and {\it
only} in this case) the real part of the localization tensor, \equ{defd}, has
a meaningful expression in terms of the one-body reduced density
matrix~\cite{rap_a23,rap118}: \[ \mbox{Re } \langle r_\alpha r_\beta
\rangle_{\rm c}\! = \! \frac{1}{2N} \!\! \int \!\! d{\bf r} d{\bf r'} ({\bf
r} - {\bf r'})_\alpha ({\bf r} - {\bf r'})_\beta | \rho^{(1)}({\bf r},{\bf
r'}) |^2 \label{dm1} , \] where single occupancy is assumed. The integral
converges whenever the density matrix vanishes fast enough at large $|{\bf r}
- {\bf r'}|$: therefore the localization tensor discriminates between
insulators and metals by measuring via \equ{dm1} the
``nearsightedness''~\cite{Kohn96} of the electron distribution. Our major,
very general, result implies that the finiteness of \equ{dm1} warrants
quantization of dc transverse conductivity.

Noninteracting electrons are kept in the quantum-Hall regime by disorder, and an
analytical implementation of the present formalism is obviously not possible. In
order to demonstrate how the theory works, I consider the academical case of a
flat substrate potential, with  noninteracting electrons in the lowest Landau
level. I show explicitly that the system is insulating, in the sense of TIS, at
complete filling, and metallic otherwise.

For complete filling ($\nu = 1$) the system is uniform with density $n_0$,
\equ{density}; the modulus of the density matrix is gauge-invariant and
equals $n_0 \exp -[({\bf r} - {\bf r'})^2 / (4 l^2) ]$.  The trace of the
localization tensor $\langle r^2 \rangle_{\rm c} = \langle x^2 \rangle_{\rm
c} + \langle y^2 \rangle_{\rm c}$ is \begin{eqnarray} \langle r^2
\rangle_{\rm c} & =& \frac{1}{2n_0} \intr r^2 | \rho^{(1)}(0,{\bf r}) |^2
\nonumber \\ & = & \pi n_0 \int_0^\infty \!\!\! dr \; r^3 {\rm
e}^{-r^2/(2l^2)} = l^2 , \label{dm2} \end{eqnarray} and therefore it equals
precisely the squared magnetic length. 

The case of $B=0$ is qualitatively different: the density matrix is
polynomial (instead of exponential) in $|{\bf r} - {\bf r'}|$, and not
nearsighted enough to make the integral in \equ{dm1} convergent. Therefore
the real part of the localization tensor is formally infinite, as expected,
while its imaginary part vanishes owing to time-reversal symmetry.  At finite
$B$ values, instead, the convergence of the real part of the tensor (hence
the insulating nature of the system) can be regarded as the cause for
quantization of the transverse conductivity.

Any single-determinant wavefunction is invariant by unitary transformations
of the occupied orbitals among themselves, and in particular by
transformations which localize the orbitals; in the general case the
localized orbitals are not eigenstates of the single-particle Hamiltonian.
The real part of the localization tensor, \equ{dm1}, provides an important
bound for such transformations~\cite{rap_a23,rap118}. Suppose one looks for
orbitals which are optimally localized in one Cartesian direction say $x$,
and delocalized along $y$. These orbitals have been called ``hermaphrodite
orbitals'' in Ref.\cite{rap118}: their quadratic spread in the $x$ direction
is minimum and equals the tensor element $\langle x^2 \rangle_{\rm c}$. 

For electrons in the lowest Landau level at complete filling, any unitary
transformation of the occupied orbitals among themselves leads to Hamiltonian
eigenstates, owing to energy degeneracy. In this case the hermaphrodite
orbitals are easily identified with the Landau-gauge
orbitals~\cite{Yoshioka}: \[ \psi_k({\bf r}) \propto \ei{k_y y} \, {\rm
e}^{-(x + k_yl^2)^2/(2l^2)} .  \label{orbl} \]In fact these orbitals are
plane-wave-like in the $y$ direction, while their quadratic spread in the $x$
direction equals precisely $\langle x^2 \rangle_{\rm c} = l^2/2$.

Next, we consider a single case study at fractional $\nu$ where the
longitudinal conductivity does not vanish and therefore---according to
TIS---the ground state is delocalized. It is expedient to switch to the
central gauge, where the single-particle orbitals are: \[ \psi_m(z) =
\frac{1}{\sqrt{2 \pi 2^m m!} \; l} \, z^m {\rm e}^{-|z|^2/4} , \] where $z =
(x - iy)/l$. Any state with fractional filling is nonuniform. A possible
state with $\nu = 1/2$ is built by occupying the odd-$m$ orbitals only, i.e.:
\begin{eqnarray} \rho^{(1)}(z,z') & = & \sum_{m=0}^{\infty} \psi_{2m+1}(z)
\psi^*_{2m+1}(z') \nonumber \\ & = & \frac{1}{2 \pi l^2} {\rm e}^{-|z|^2/4}
{\rm e}^{-|z'|^2/4} \sinh(z z'^*/2) .  \label{rho} \end{eqnarray} This
density matrix is {\it not} nearsighted: taking for instance $z' = -z$ we
have \[ \rho^{(1)}(z,-z) = - \frac{1}{2 \pi l^2} {\rm e}^{-|z|^2/2 }
\sinh(|z|^2/2), \] which clearly does not vanish at large $|z|$. The integral in
\equ{dm1} is positively divergent, providing a formally infinite real part of
the localization tensor, as expected. Because of the above general
considerations, the corresponding imaginary part is ill-defined and the
transverse conductivity is not quantized.

In conclusion, I have shown quite generally that the TIS localization
tensor~\cite{rap107,Souza00,rap_a23}---besides discriminating between insulators
and metals on the basis of longitudinal conductivity---also yields very directly
the transverse dissipationless dc conductivity in the insulating case, as e.g.
in a quantum-Hall fluid. It is enough to inspect electron localization in order to
predict whether the dc transverse conductivity is quantized. The localization
tensor is a pure ground-state property and has a geometric nature: it coincides
in fact with the quantum metric-and-curvature tensor of Provost and
Vallee~\cite{Provost80} (divided by $N$), \eqs{PV}{discre}. Both the real and
the imaginary parts of the TIS localization tensor carry an outstanding physical
meaning.

Work partly supported by ONR through grant N00014-03-1-0570.


\begin{thebibliography}{10}

\bibitem{Kohn64}
{ W. Kohn, Phys. Rev. {\bf 133}, {A171} (1964)}.

\bibitem{Kohn68}
{ W. Kohn, in {\it Many--Body Physics}, edited by C. DeWitt and R. Balian
  (Gordon and Breach, New York, 1968), p. 351}.

\bibitem{rap107}
{ R. Resta and S. Sorella, Phys. Rev. Lett. {\bf 82}, 370 (1999)}.

\bibitem{modern}
{ R. D. King-Smith and D. Vanderbilt, Phys.\ Rev.\ B {\bf 47}, 1651 (1993); R.
  Resta, Rev. Mod. Phys. {\bf 66}, 899 (1994)}.

\bibitem{rap100}
{ R. Resta, Phys. Rev. Lett. {\bf 80}, 1800 (1998)}.

\bibitem{Martin22}
{ R. M. Martin, {\it Electronc Structure: Basic Theory and Practical Methods}
  (Cambridge University Press, 2004), Chap. 22}.

\bibitem{Souza00}
{ I. Souza, T. Wilkens, and R. M. Martin, Phys. Rev. B {\bf 62}, 1666 (2000)}.

\bibitem{rap_a23}
{ R. Resta, J. Phys.: Condens. Matter {\bf 14}, R625 (2002)}.

\bibitem{Niu85}
{ Q. Niu, D. J. Thouless, and Y. S. Wu, Phys. Rev. B {\bf 31}, 3372 (1985)}.

\bibitem{Yoshioka}
{ D. Yoshioka, {\it The Quantum Hall Effect} (Springer, Berlin, 2002)}.

\bibitem{Akkermans97}
{ E. Akkermans, J. Math. Phys. {\bf 38}, 1781 (1997)}.

\bibitem{Provost80}
{ J. P. Provost and G. Vallee, Commun. Math Phys. {\bf 76}, 289 (1980)}.

\bibitem{Aligia}
{ A. A. Aligia and E. R. Gagliano, Physica C {\bf 304}, 29 (1998); A. A.
  Aligia, Europhys. Lett. {\bf 45}, 411 (1999)}.

\bibitem{rap_a20}
{ R. Resta, J. Phys.: Condens. Matter {\bf 12}, R107 (2000)}.

\bibitem{Kohmoto85}
{ M. Kohmoto, Ann. Phys. {\bf 160}, 343 (1985)}.

\bibitem{Thouless}
{ D. J. Thouless, {\it Topological Quantum Numbers in Nonrelativistic Physics}
  (World Scientific, Singapore, 1998)}.

\bibitem{Haldane88}
{ F. D. M. Haldane, Phys. Rev. Lett. {\bf 61}, 2015 (1988)}.

\bibitem{Stone3}
{ M. Stone, {\it Quantum Hall Effect} (World Scientific, Singapore, 1992),
  Chap. 3}.

\bibitem{rap118}
{ C. Sgiarovello, M. Peressi, and R. Resta, Phys. Rev. {\bf 64}, 115202
  (2001)}.

\bibitem{Kohn96}
{ W. Kohn, Phys. Rev. Lett. {\bf 76}, 3168 (1996)}.

\end{thebibliography}
\end{document}